\documentclass[12pt]{article}
 \pdfoutput=1
 \usepackage{graphicx}
  \usepackage{epsfig}
  \usepackage{amsmath}
  \usepackage{amssymb}
  \usepackage{color}
  \newlength{\abstractwidth}
  \newcommand{\be}{\begin{equation}}
  \newcommand{\ee}{\end{equation}}
  
  \renewcommand{\title}[1]{\vbox{\center\bf{\Large{#1}}}\vspace{5mm}}
  \renewcommand{\author}[1]{\vbox{\center#1}\vspace{5mm}}
  \newcommand{\address}[1]{\vbox{\center\em#1}}
  \newcommand{\email}[1]{\vbox{\center\tt#1}\vspace{5mm}}

\begin{document}

\begin{titlepage}
\rightline{SU-ITP-13/24}
\begin{center}
\hfill \\
\hfill \\
\vskip 1cm

\title{Multiple Shocks}

\author{Stephen H. Shenker and Douglas Stanford}

\address{
 Stanford Institute for Theoretical Physics {\it and} \\
 Department of Physics, Stanford University \\
 Stanford, CA 94305 USA
}

\email{sshenker@stanford.edu,
salguod@stanford.edu}

\end{center}
  
  \begin{abstract}
 Using gauge/gravity duality, we explore a class of states of two CFTs with a large degree of entanglement, but with very weak local two-sided correlation. These states are constructed by perturbing the thermofield double state with thermal-scale operators that are local at different times. Acting on the dual black hole geometry, these perturbations create an intersecting network of shock waves, supporting a very long wormhole.   Chaotic CFT dynamics and the associated fast scrambling time play an essential role in determining the qualitative features of the resulting geometries.

  \end{abstract}

  \end{titlepage}

\tableofcontents

\baselineskip=17.63pt

\section{Introduction}

The firewall \cite{AMPS} controversy has highlighted the conflict between the special local entanglements required for smooth geometry and the randomness of typical states.    Aspects of this tension become especially clear in the two sided  black hole \cite{Maldacena:2001kr,israel} context, as Van Raamsdonk  has emphasized.   The two sided eternal AdS Schwarzschild black hole is dual to two copies of a CFT, $L$ (left) and $R$ (right), in the thermofield double state
\be
|TFD\rangle = \frac{1}{Z^{1/2}}\sum_n e^{-\beta E_n/2}|n\rangle_L|n\rangle_R.
\ee
The particular $LR$ entanglement in this state is highly atypical, as local subsystems of $L$ are entangled with local subsystems of $R$. This structure is closely related to the smooth geometry of the eternal black hole. The primary goal of this paper is to explore how geometry can respond to operations that delocalize the entanglement.

Van Raamsdonk \cite{VanRaamsdonk:2013sza} pointed out that a random unitary transformation applied to the left handed CFT leaves the density matrix describing right handed CFT observables unchanged, but will change the relation between degrees of freedom on both sides and hence the geometry behind the horizon. Certain unitaries correspond to local operators, which can create a pulse of radiation propagating just behind the horizon \cite{VanRaamsdonk}.

We examined this situation in detail in our study of scrambling \cite{SS}.  We showed that a local operator on the left hand boundary that only injects one thermal quantum worth of energy, if applied early enough, scrambles the left hand Hilbert space and disrupts the special local entanglement.   This happens when the time since the perturbation, $t_w$, is of order the fast scrambling time \cite{Sekino:2008he,Hayden:2007cs}\footnote{The importance of this timescale to black hole physics was pointed out in earlier work, including \cite{Schoutens:1993hu}.}
\be
t_* = \frac{\beta}{2\pi} \log S
\ee
where $S$ is the black hole entropy and $\beta$ is the inverse temperature.  From the bulk point of view, the perturbation sourced at an early time (large $t_w$) is highly boosted relative to the $t=0$  frame, creating a shock wave, as illustrated in the right panel of Fig.~\ref{figone}.  This shock disrupts the Ryu Takayanagi surface \cite{Ryu:2006bv,Hubeny:2007xt} passing through the wormhole \cite{Morrison:2012iz,Hartman:2013qma}. The area of this surface is used to calculate the mutual information $I(A,B)$ that diagnoses the special entanglement between local subsystems $A\subset L$, $B\subset R$ of the two CFTs.   For subsystems smaller than half, one finds that the leading contribution to $I$ drops to zero when $t_w \sim t_*$.  

The two point correlation function $\langle \varphi_L(t)\varphi_R(t)\rangle$, with operators at equal Killing time on opposite sides, also diagnoses the relation between degrees of freedom and should become small if $|t - t_w|$ is of order the scrambling time.  In the bulk it is related to geodesics and hence probes the geometry \cite{Balasubramanian:1999zv, Louko:2000tp, Kraus:2002iv, Fidkowski:2003nf,Kaplan:2004qe, Festuccia:2005pi}.  Using (2+1) Einstein gravity and ignoring nonlinear effects, the correlation function was computed in \cite{SS}, using the length of the geodesic connecting the correlated points. Roughly, the result decreases like a power of $1/\left(1 + e^{2\pi(|t-t_w|-t_*)/\beta}\right)$. The fact that this expression depends only on $(t - t_w)$ is a consequence of the boost symmetry of the eternal black hole. It is clear that, for any choice of $t_w$, there is a time $t \sim t_w$ at which the correlator $\langle \varphi_L(t)\varphi_R(t)\rangle$ is order one.

As pointed out in \cite{SS}, when $|t -t_w|$ is large, the relative boost between the geodesic and the shock wave is very large.  This makes likely the possibility that nonlinear corrections to the correlation function result are important.   We are currently exploring these effects but in this paper we will ignore them.   We hope the Einstein gravity results will be a useful guide to the important phenomena.   In any event they should serve as a lower bound to the strength of these effects.

Marolf and Polchinski \cite{Marolf:2013dba} analyzed the behavior of truly typical two sided states where the average energy of the total Hamiltonian $H_L + H_R$ is fixed.  Using the Eigenvector Thermalization Hypothesis \cite{ETH}, they showed that the two point correlator between local operators on the two sides is typically $\sim e^{-S}$, and is never larger than $\sim e^{-S/2}$, for any choice of times for the two operators. This is in contrast with the behavior of correlators in the shock wave geometry discussed above. Marolf and Polchinski interpreted their result as evidence for a ``non geometrical" connection between the two sides.

The work of Maldacena and Susskind \cite{er=epr} suggests a different potential interpretation. These authors considered the time evolution of the thermofield double state\footnote{Here, we mean time evolution with $H_L + H_R$.} as a family of states in which the local entanglements present in $|TFD\rangle$ are disturbed. At late times, two-sided correlations become small because of the increasing length of the geodesic threading the wormhole. This suggests that the behavior found in \cite{Marolf:2013dba} could be consistent with a smooth but very long wormhole linking the two sides.

In fact, very little is known about more general states. To this end, we explore in \S\ref{wormholes} a class of geometries obtained by perturbing the left side of the thermofield double state with a string of unitary local operators with order-one energy,
\be
W_n(t_n)...W_1(t_1)|TFD\rangle.\label{hello}
\ee
If the time separations are sufficiently large, the boosting effect described above means that these states are dual to geometries with $n$ shock waves. We will outline an iterative procedure that builds the geometry one shock wave at a time. Using this method, we will explore a small part of the diverse class of metrics dual to states of this form. If the time separations and/or the number of shocks is large, one finds that the wormhole connecting the two asymptotic regions becomes very long in all boost frames, indicating weak local correlation between the two boundaries at all times. 

The timescale $t_*$ plays a central role in the construction, indicating that the geometry is sensitive to chaotic dynamics in the CFT. The application of a $W$ operator creates a short-distance disturbance in the CFT. The application of a second, at time separation greater than $t_*$, creates a second disturbance and erases the first. This manifestation of scrambling is represented in the bulk by the second shock wave pushing the first off the AdS boundary and onto the singularity.

The states (\ref{hello}) and their bulk duals provide examples of how Einstein gravity can accommodate weak two-sided correlations, but they are not typical in the Hilbert space. This is for multiple reasons. First, the $W$ operators inject some energy into one of the CFTs, making the energy statistics not precisely thermal. Second, the operators leave a distinguished time $t_n$ at which a local perturbation is detectable in the left CFT. In order to make states with weak two-sided correlation, we pay the price of an atypical $\rho_L$.

In general, the duals to (\ref{hello}) are geometrical, but they are not drama-free. In particular, by boosting the geometry one way or another, one can always find a frame in which an infalling observer collides with a high energy shock very near the horizon. In \S\ref{ensembles}, we will emphasize that the class of truly typical states should be invariant under such boosts. This constrains the possible form of a smooth geometrical dual to a typical state.

We will conclude in \S\ref{discussion}. Certain technical details of the shock wave construction are recorded in two appendices.

AdS/CFT applications of wide wormholes have previously been discussed in \cite{Bak:2007jm}. In \cite{er=epr}, it was noted that adding matter at the boundaries of the eternal black hole would make a wide wormhole describing less than maximal entanglement. Our examples are similar, but we add a small amount of matter, relying on the effect of \cite{SS} to amplify the perturbation, and leaving the total entanglement near maximal. The length of the resulting wormhole is related to the absence of local two-sided entanglement \cite{Hartman:2013qma}. The paper \cite{Susskind:2013aaa} contains further discussion of the connection between chaos and geometry described here. 

\section{Wormholes built from shock waves}\label{wormholes}
\subsection{One shock}
Let us begin by reviewing the geometrical dual to a single perturbation of the thermofield double \cite{SS}. We consider a CFT state of the form
\be
\label{oneW}
W(t_1)|TFD\rangle,
\ee
where the operator $W$ acts unitarily on the left CFT and raises the energy by an amount $E$. The scale $E$ is assumed to be of order the temperature of the black hole, much smaller than the mass $M$.\footnote{For a large AdS black hole dual to a state with temperature of order the AdS scale, we have $E\sim 1$ in AdS units, while $M \sim 1/G_N$, which is proportional to $N^2$ in the large-$N$ gauge theory.} To keep the bulk solutions as simple as possible, we will assume that $W$ acts in an approximately spherically symmetric manner. We will also assume that $W$ is built from local operators in such a way that it acts near the boundary of the bulk AdS space.

One can think about the expression (\ref{oneW}) in different ways. One option would be to understand it as a thermofield double state that was actively perturbed by a source at time $t_1$; the $W$ operator would then be time-ordered relative to other operators in an expectation value. Another option is to understand it as the state of a system evolving with a strictly time-independent Hamiltonian. We will occasionally use language appropriate to the first interpretation, but where it makes a difference (i.e. for expectation values involving operators before $t_1$) we will stick to the second, ordering the $W$ operator immediately after the state vector.
\begin{figure}[ht]
\begin{center}
\includegraphics[scale = 0.75]{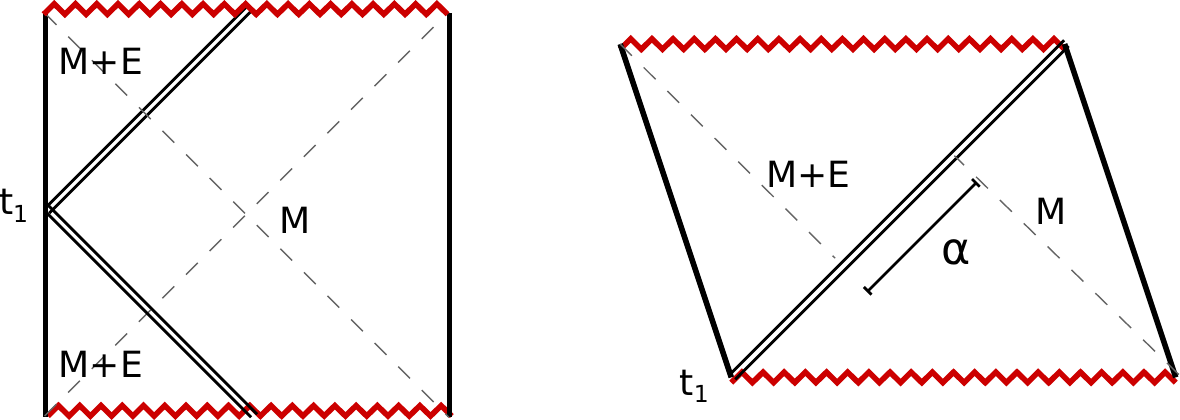}
\caption{The geometry dual to Eq.~(\ref{oneW}) consists of a perturbation that emerges from the past horizon and falls through the future horizon (left). If $t_1$ is sufficiently early, the boost relative to the $t = 0$ slice generates backreaction in that frame (right). Note that the horizons no longer meet.} \label{figone}
\end{center}
\end{figure}

With this understanding, the bulk dual to the state (\ref{oneW}) consists of a perturbation that emerges from the past horizon of the black hole, approaches the boundary at time $t_1$, and then falls through the future horizon, as shown in the left panel of Fig.~\ref{figone}. Since the energy scale of the perturbation is order one, backreaction on the metric is negligible. However, if we increase the Killing time $t_1$, the perturbation is boosted relative to the original frame, and the energy relative to the horizontal $t = 0$ surface increases as\footnote{In our conventions, the Killing time $t$ increases downwards on the left boundary.}
\be
E_p^{(t = 0)} \sim E e^{2\pi t_1/\beta},
\ee
where $\ell$ is the AdS radius, and $\beta$ is the inverse temperature of the black hole. Once $t_1 \sim t_*$, backreaction must be included. The resulting geometry is sketched in the right panel.\footnote{Notice that we have represented the matter as a thin-wall null shell. Physical perturbations will have some spatial width, and they might follow massive trajectories. However, because of the highly boosted kinematics that we will consider in this paper, it will be permissible to treat all matter in this way.} Details of the shock wave metric are given in \cite{SS}, following earlier work by \cite{Dray:1984ha,Hotta:1992qy, Sfetsos:1994xa, Cai:1999dz}. For the remainder of this section, we will work in the (2+1) dimensional setting of the BTZ black hole. This is for technical convenience; the essential features generalize to higher dimensions. For small $E$ and large $t_1$, a good approximation to this metric consists of two pieces of the same BTZ geometry, glued together across the $u = 0$ surface, with a null shift in the $v$ coordinate by amount
\be
\alpha = \frac{E}{4M}e^{2\pi t_1/\beta} \sim e^{2\pi(t_1 - t_*)/\beta}.
\ee
Here, we are using Kruskal coordinates for each of the patches, with metric
\be
ds^2 = \frac{-4\ell^2dudv + R^2(1-uv)^2d\phi^2}{(1+uv)^2}.\label{kruskal}
\ee

\subsection{Two shocks}
Next, we consider a state of the form
\be
W(t_2)W(t_1)|TFD\rangle.
\ee
To construct the bulk dual, we simply need to act with $W(t_2)$ on the single-shock geometry constructed above. In order to do this, it is helpful to generalize our problem slightly, and understand how to construct the bulk dual to a state
\be
W(t)|\Phi\rangle,
\ee
assuming that we already know the geometry for $|\Phi\rangle$. In general, the prescription is as follows: we start with the geometry for $|\Phi\rangle$ and select a bulk Cauchy surface that touches the left boundary at time $t$. We record the data on that surface, add the perturbation corresponding to $W(t)$ near the boundary, and evolve the new data forwards and backwards. 

\begin{figure}[ht]
\begin{center}
\includegraphics[scale = 0.65]{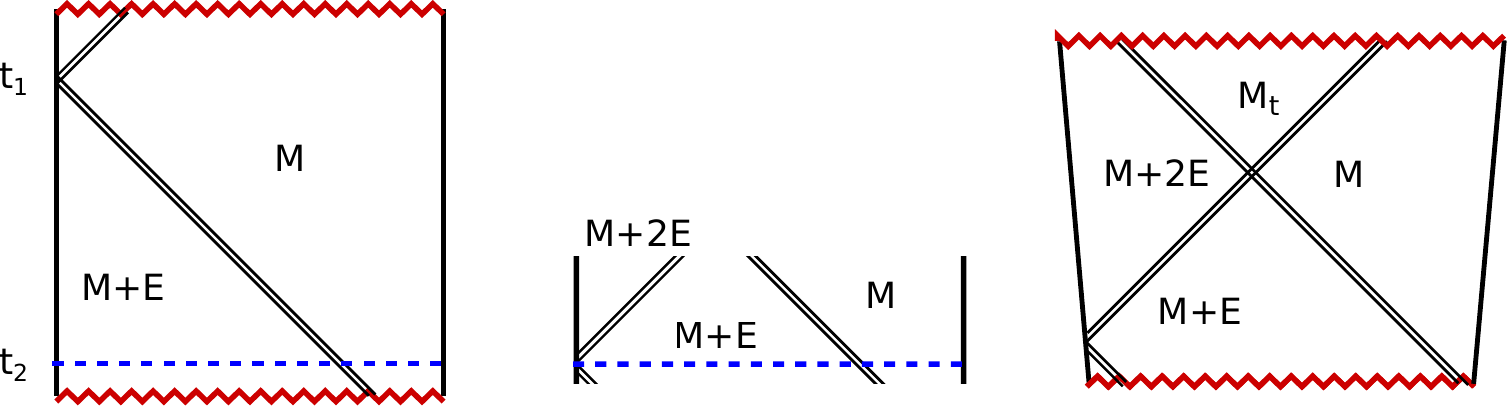}
\caption{The dual to a two-$W$ state is constructed from the one-$W$ state by adding a perturbation near the boundary at time $t_2$ and then evolving forwards and backwards.}\label{figtwo}
\end{center}
\end{figure}

In Fig.~\ref{figtwo}, we use the above procedure to build the two-$W$ geometry. The left panel represents the state $W(t_1)|TFD\rangle$, and the dashed blue line is the Cauchy surface that touches the left boundary at time $t_2$. We add the second perturbation and evolve forwards and backwards in time, producing the geometry shown on the right.

We can understand this prescription in terms of the ``folded'' bulk geometries discussed in \cite{HMPS}. The two-shock geometry corresponds to a folded bulk with three sheets. On the first sheet, we evolve from $-\infty$ to $t_1$. On the second sheet (a portion of the left panel of Fig.~\ref{figtwo}), we add a perturbation at $t_1$ and evolve backwards in time from $t_1$ to $t_2$. On the final sheet (a portion of the right panel of Fig.~\ref{figtwo}), we add a perturbation at $t_2$ and evolve forwards to $+\infty$. Our prescription to order the $W$ operators immediately after the state means that we focus on the final fold of the bulk, extending it in time from $-\infty$ to $+\infty$, however we use each of the sheets in our iterative construction procedure.

It is clear from the figure that the two shells collide on the final sheet. Our assumptions of spherical symmetry and thin walls make it possible to construct the full geometry by pasting together AdS-Schwarzschild geometries with different masses. There are two conditions: first, we require $r$, the size of the sphere, to be continuous at the join. Second, we have the DTR regularity condition \cite{Dray:1985yt,redmount,Poisson:1990eh}
\be
f_{t}(r)f_{b}(r) = f_{l}(r)f_{r}(r),\label{DTR}
\ee
where $t,b,l,r$ refer to the top, bottom, left and right quadrants, and $f$ is the factor in the metric $ds^2 = -f dt^2 + f^{-1} dr^2+r^2d\Omega^2$. Explicitly, for the 2+1 dimensional BTZ case, $f(r) = r^2 - 8G_NM\ell^2$, where $M$ is the mass of the black hole and $\ell$ is the AdS length. The DTR condition becomes
\be
\big[r^2 - 8G_N M_{t}\ell^2\big]\big[r^2 - 8G_N (M+E)\ell^2\big] = \big[r^2 - 8G_N (M+2E)\ell^2\big]\big[r^2 - 8G_NM\ell^2\big].\label{condition}
\ee
If the collision takes place at large $r$, the evolution is nearly linear and this equation implements conservation of energy of the shells. However, even beyond the linear regime, the equation plays a similar role, fixing the mass $M_t$ of the Schwarzschild solution in the post-collision region in terms of the other masses and $r$, the radius of the collision. In turn, $r$ is set by the time difference $(t_2 - t_1)$. To find the precise relation, it is simplest to use Kruskal coordinates. By matching the size of the $S^1$ in the two coordinate systems, we find that $r$ is determined by $u$ and $v$ as
\be
\frac{r}{R} = \frac{1 - uv}{1+uv},\label{ruv}
\ee
where the radius of the horizon, $R$, is determined by $R^2 = 8G_N M \ell^2$, with $M$ is the mass of the black hole and $\ell$ the AdS length. The $u$ and $v$ coordinates are conserved, respectively, by right-moving and left-moving radial null trajectories. Using the Kruskal conventions in \cite{SS}, we can determine the value of $u$ or $v$ using the time coordinate at which the trajectory hits the left boundary:
\be
u = e^{-R t/\ell^2}, \hspace{20pt} v = -e^{R t/\ell^2}.\label{uv}
\ee
In particular, in the Kruskal system of the bottom quadrant, the $v$ coordinate of the left-moving shock is $-e^{R_bt_1/\ell^2}$, while the $u$ coordinate of the right-moving shock is $e^{-R_bt_2/\ell^2}$.\footnote{$R_b$ is the BTZ radius in the lower quadrant, defined by $R_b^2 = 8 G_N (M + E)\ell^2$.} This determines the $r$ value of their collision as
\be
\frac{r}{R_b} = \frac{1 + e^{R_b(t_1 - t_2)/\ell^2}}{1 - e^{R_b(t_1 - t_2)/\ell^2}}\label{r}.
\ee
Plugging this value of $r$ into Eq.~(\ref{condition}), we find
\be
M_t  = M+E + \frac{E^2}{M+E}\sinh^2 \frac{R_b(t_2-t_1)}{2\ell^2}.\label{mass}
\ee
The final, exponentially growing term begins to dominate the first term when $(t_2 - t_1) \approx 2t_*$.

\begin{figure}[ht]
\begin{center}
\includegraphics[scale = 0.65]{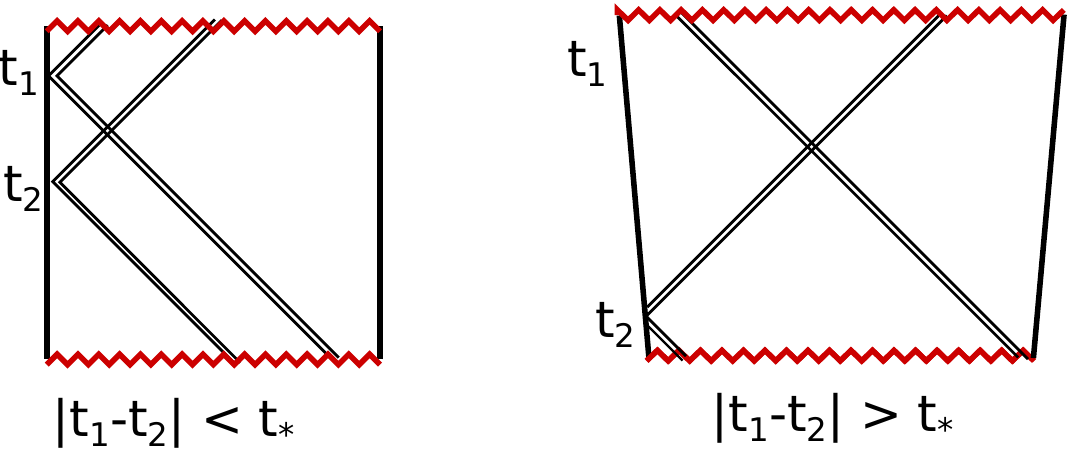}
\caption{As $t_2$ shifts earlier, the time at which the original shock reaches the boundary shifts later, eventually moving onto the singularity (right).}\label{fig8}
\end{center}
\end{figure}

Given that a $W(t)$ operator creates a perturbation in the UV at time $t$, one might have expected a two-$W$ state to have perturbations near the boundary both at $t_1$ and at $t_2$. In fact, if the time difference is greater than scrambling, this is not the case. In the bulk, we can understand this by going back to the  left panel of Fig.~\ref{figtwo}. In this one-$W$ state, the $W(t_1)$ perturbation approaches the boundary at time $t_1$, but at much earlier times it is very close to the horizon. If we add the second perturbation $W(t_2)$ sufficiently early, then the outward jump of the horizon due to the increase in mass will be enough to capture the first shock, as shown in the right panel of Fig.~\ref{fig8}.

To analyze this effect in detail, it is again helpful to use Kruskal coordinates. The key is to determine the $v$ coordinate of the trajectory of the $W(t_1)$ shell in the Kruskal system of the left quadrant. If $v$ is negative, then the shell hits the boundary at time $e^{Rt/\ell^2} = -v$. If $v$ is positive, then the shell runs from singularity to singularity. To find the $v$ coordinate, we can use Eq.~(\ref{ruv}), plugging in the $r$ coordinate in Eq.~(\ref{r}), and the $u$ coordinate in the left Kruskal system $e^{-R_l t_2/\ell^2} \approx e^{-R t_2/\ell^2}$. We find
\be
v \approx -e^{R t_1/\ell^2} + \frac{E}{4M}e^{R t_2/\ell^2}.\label{v}
\ee 
The coordinate becomes positive, indicating that the shock wave has moved off the left boundary and onto the singularity, when $(t_2 - t_1) \approx t_*$.

The presence of the timescale $t_*$ suggests that we interpret the ``capture'' of the first perturbation in terms of scrambling. Indeed, the state $W(t_1)|TFD\rangle$ is carefully tuned to produce an atypical perturbation in the UV at time $t_1$. If we additionally perturb this state by acting with $W(t_2)$ a scrambling time before $t_1$, this delicate tuning is upset, and the perturbation at $t_1$ fails to materialize.

We can also think about this effect in terms of the square of the commutator
\be
\langle TFD| [W_1(t_1),W_2(t_2)]^\dagger [W_1(t_1),W_2(t_2)]|TFD\rangle.\label{comm}
\ee
Expanding this out, we find two terms that each give a numerical contribution of one, minus two terms involving the overlap of $W_1(t_1)W_2(t_2)|TFD\rangle$ and $W_2(t_2)W_1(t_1)|TFD\rangle$. According to the bulk solution just described, the overlap of these states should be small if the time separation is greater than $t_*$, indicating that (\ref{comm}) becomes approximately equal to two once $|t_1-t_2|\sim t_*$. This large commutator is a sharp diagnostic of chaos: perturbing one quantum perturbs all quanta a scrambling time later \cite{AMPSS}.

\subsection{Many shocks}
A general geometry built from spherical shock waves can be analyzed in terms of a sequence of two-shock collisions. This means that the matching conditions discussed above, together with the recursive procedure for adding a $W$ perturbation, allow us to construct the dual to arbitrary states of the form
\be
W_n(t_n)...W_1(t_1)|TFD\rangle.
\ee
By varying the times $t_1,...,t_n$, one finds a very wide array of possible metrics. We will focus on a particular slice through the space of these states, in which all even-numbered times are equal to $t_w$, and all odd-numbered times are equal to $-t_w$.

\begin{figure}[ht]
\begin{center}
\includegraphics[scale = 0.4]{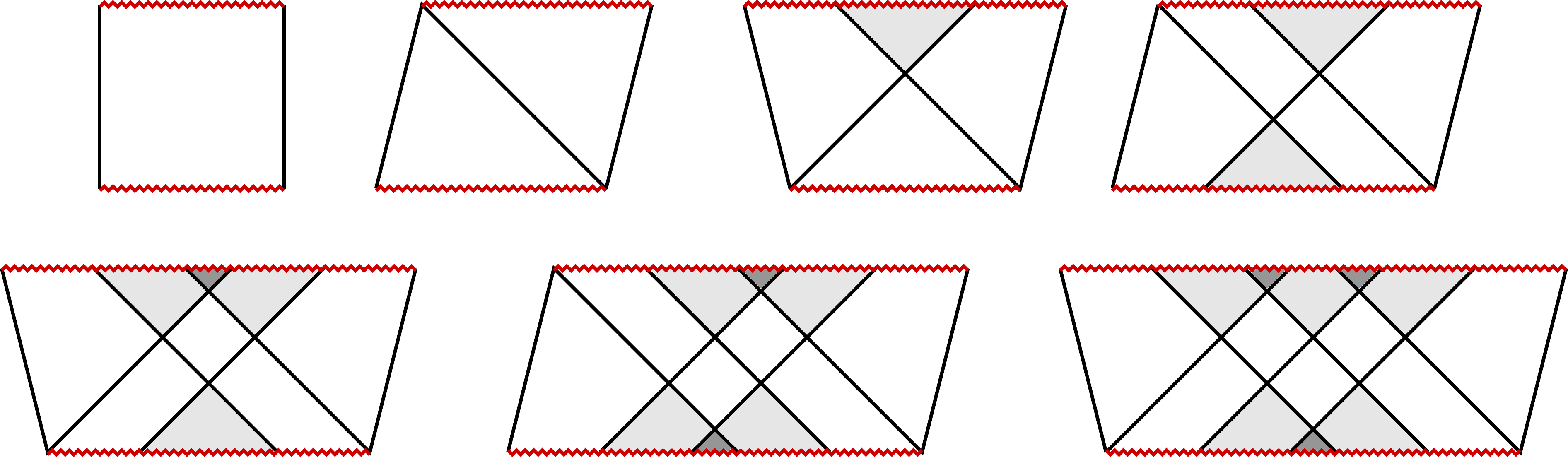}
\caption{The thermofield double and the first six multi-$W$ states are drawn. In each case, the next geometry is obtained from the previous by adding a shock either from the top left or bottom left corner. The gray regions are sensitive to the details of a collision, but the white regions are not. Using the time-folded bulk of \cite{HMPS}, these states can be combined as different sheets of an ``accordion'' geometry.}\label{figiteration}
\end{center}
\end{figure}

We will also assume that the asymptotic energy of each shock, $E$, is very small compared to the unperturbed mass $M$. The large-$N$ limit in the gauge theory allows us to take $E/M\rightarrow 0$ and $t_w\rightarrow \infty$, with 
\be
\alpha = \frac{E}{4M}e^{2\pi t_w/\beta}
\ee
held fixed. In this limit, the iterative construction process described above becomes rather straightforward: we alternately add shocks traveling forwards in time from the bottom left corner, and backwards in time from the top left. The associated null shifts, which alternate in the $u$ and $v$ directions, have the effect of extending the wormhole to the left, as illustrated in Fig.~\ref{figiteration}.

Because of the null shifts, all but one of the shock waves run from singularity to singularity. Still, the leftmost one touches the boundary at time $\pm t_w$,\footnote{Here, we are backing off the limit $t_w\rightarrow \infty$.} making this time locally distinguished in the CFT. One can also consider bulk solutions with the property that all shocks run from singularity to singularity, leaving no locally distinguished time. At the level of the bulk theory, there is nothing wrong with these geometries. However, unlike the multi-$W$ states described in this paper, we are not sure how or whether they can be constructed in the CFT.

Our assumption that the $\{t_i\}$ are equal in magnitude and alternating in sign means that the interior region of the resulting wormhole has a discrete translation symmetry. This is simplest to understand if we consider building a geometry from an infinite sequence of shocks. After $k$ steps in the iterative procedure, the geometry to the left of all $k$ shocks will be unperturbed AdS-Schwarzschild. The geometry that gets built in that region by the remaining (infinite) collection of shocks is therefore the same as the geometry to the left of the first $(k+2)$ shocks.\footnote{Notice that at finite $E$, this symmetry would be broken by a smoothly varying mass profile in the wormhole, increasing from right to left. If we relax the assumption of equal times, this translation invariance would also be broken by the fact that different $W$ operators source shocks of varying strength.}

Using this translation invariance, we can understand the full geometry of the wormhole by studying a ``unit cell,'' for which the geometry depends on $\alpha$ but not $n$. Let us begin by computing the length of the wormhole, i.e. the regularized length of the shortest geodesic that passes from the left boundary to the right. Up to an $n$-independent deficit, this is simply $n$ times the length across the central layer of a unit cell.
\begin{figure}[ht]
\begin{center}
\includegraphics[scale = 0.65]{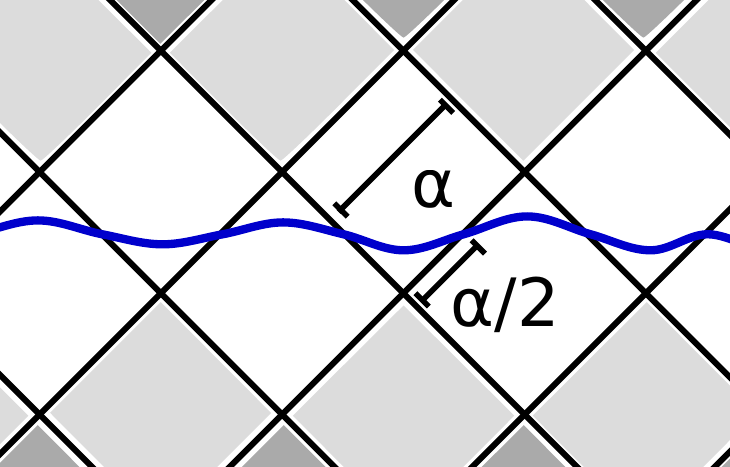}
\caption{A geodesic passes across a portion of the wormhole. It intersects the null boundaries of the central regions halfway across their width.}\label{figeight}
\end{center}
\end{figure}
The portion of the geodesic that passes through this unit cell (see Fig.~\ref{figeight}) is a geodesic in the BTZ geometry passing from Kruskal coordinates $(u=0,\alpha/2)$ to $(u=\alpha/2,v=0)$. The length of such a geodesic is $\ell\cosh^{-1}(1 + \alpha^2/2)$. Thus the regularized length across the entire wormhole is
\be
\frac{L}{\ell} = n \cosh^{-1}\left(1+\frac{\alpha^2}{2}\right) + O(n^{0}).
\ee
This function interpolates between $n\alpha$ for small $\alpha$ and $2n\log \alpha$ for large $\alpha$. We can make this length large, and in particular greater than $S$, by making $\alpha$ and/or $n$ large. Such wormhole geometries therefore describe CFT states with very weak local correlation $\sim e^{-(const.) L}$ between the two sides. Note, however, that if we make $L\sim S$ by fixing $\alpha$ and taking $n \sim S$, then the mass of the left black hole will be larger than that of the right by an amount $\delta M \sim S E \sim M$. Instead, we could fix $n$ and take the time differences to be of order $S$. In this case, the energies of the shocks are extremely high $\sim e^{S}$, and the geometrical computation of the correlator is completely out of control. We interpret the geodesic estimate as an upper bound on the true correlator.

\begin{figure}[ht]
\begin{center}
\includegraphics[scale = 0.5]{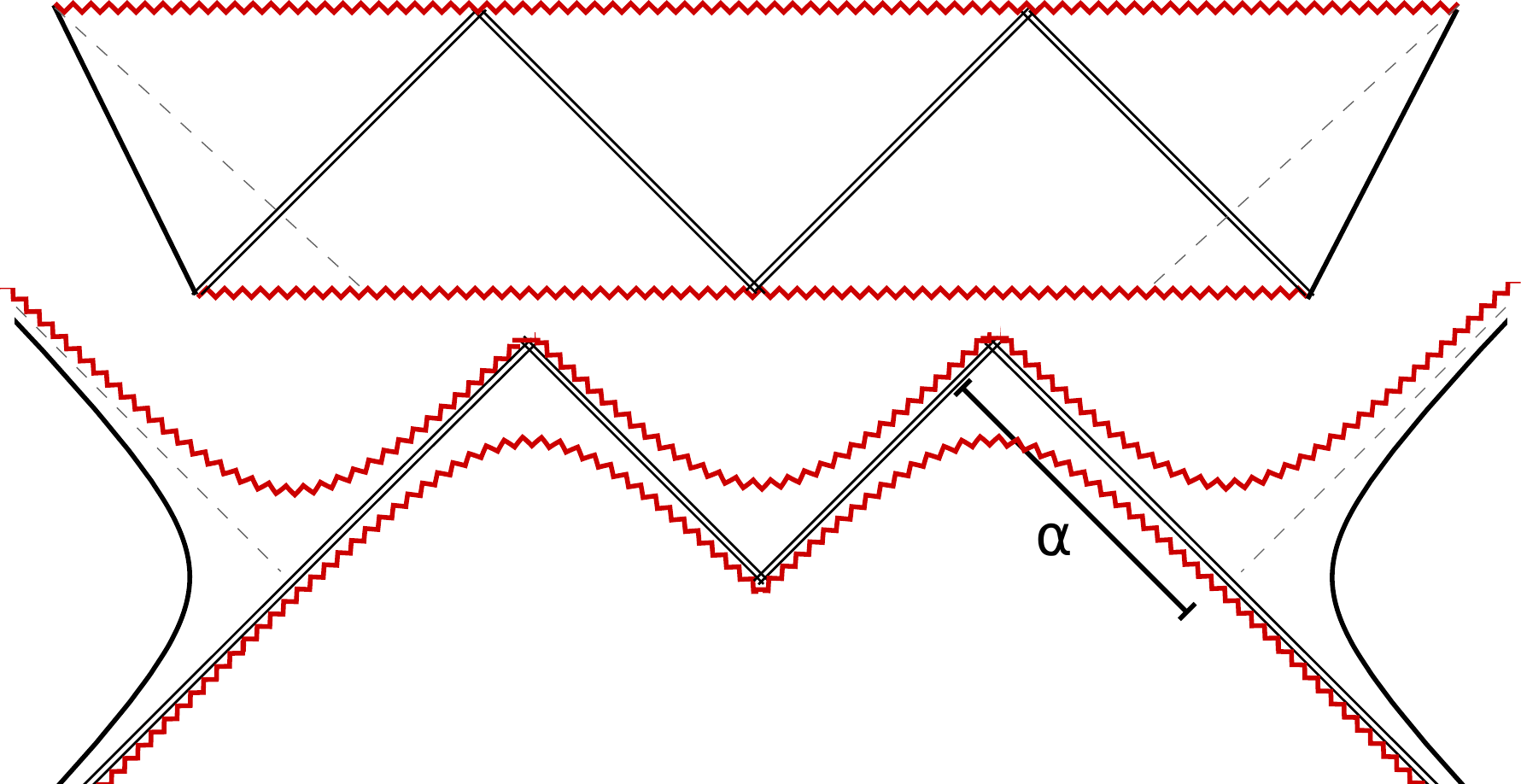}
\caption{The large-$\alpha$ four-$W$ geometry is shown. Notice that the post-collision regions are small and isolated near the singularities. The Kruskal diagram at the bottom emphasizes the kinkiness of the geometry.}\label{figfour}
\end{center}
\end{figure}

Having computed the length, we would like to understand the qualitative shape of the unit cell as a function of $\alpha$. First, let us consider the case in which $\alpha$ is large compared to one. The construction of the geometry is very simple in this limit, because the post-collision regions are pushed near the singularities, and almost none of the geometry is affected by the details of the collisions. This should be clear from the large-$\alpha$ four-$W$ geometry shown in Fig.~\ref{figfour}.

For intermediate values of $\alpha\lesssim 1$, we have geometries similar to those in Fig.~\ref{figfour}. The central white region is unaffected by details of the collisions, but the Mandelstam $s$ invariant in each collision is of order $\alpha^2 M^2$, and the shaded regions will be sensitive to string and Planck scale physics.

For small values of $\alpha$ (with $\alpha n$ fixed) it is natural to guess that the large kinks of size $\alpha$ in Fig.~\ref{figfour} will be smoothed out,\footnote{We are grateful to Raphael Bousso for making this suggestion.} allowing an analysis in terms of an averaged stress tensor. For most values of $\alpha \ll 1$, inelastic stringy effects, proportional to $G_N \alpha^2 M^2 \ell_s^2/\ell^{D-2}$ \cite{veneziano}, will be important in determining the form of this stress energy. As an example, though, we will work out the geometry appropriate for the case in which $\alpha$ is small enough that we can ignore these effects\footnote{We need $\alpha$ small enough that the probability of oscillator excitation per collision, $G_N \alpha^2 M^2 \ell_s^2/\ell^{D-2}$, times the number of collisions, $1/\alpha$, is small. Roughly, we support the wormhole with a large number of relatively soft quanta, with boost factor $e^{2\pi t_w/\beta}$ of order $\ell^2/\ell_s^2$. The mild boost means that doubling the mass of the left black hole only leads to a wormhole of length $\ell^3/\ell_s^2$.} Thus, we look for a solution to Einstein's equations with radial null matter moving in both directions, and with translation symmetry plus spherical symmetry.\footnote{In a realistic setting, the shocks won't be exactly spherically symmetric. Suppose we build each shell as a sum of particles localized on the $S^1$. After a collision, these can be deflected by an angle $\sim \alpha$ \cite{veneziano}. Each experiences $\sim 1/\alpha$ collisions before hitting the singularity, but if the initial inhomogeneity is small, deflections will tend to cancel, and the total effect will remain small.}

Specifically, we make an ansatz
\be
ds^2 = -\ell^2d\tau^2 + h(\tau)^2dx^2 + g(\tau)^2 d\phi^2
\ee
and compute the stress tensor implied Einstein's equations. In order for $T_{\phi,\phi}$ to be pure cosmological constant, $h(\tau)$ must be proportional to $\cos\tau$. In order for $T_{\tau,\tau}$ and $T_{x,x}$ to be pure cosmological constant plus traceless matter, we find an equation for $g$. By requiring that the solution be differentiable at $\tau = 0$, we find that the metric is uniquely determined (up to the scales $\ell$ and $R$, which we now restore) as:
\begin{align}
ds^2 &= -\ell^2 d\tau^2 + \ell^2\cos^2\tau dx^2 + g(\tau)^2d\phi^2 \label{metric}\\
\frac{g(\tau)}{R}& = 1 - \sin\tau \ \log\frac{1+\sin\tau}{\cos\tau}.\notag
\end{align}
\begin{figure}[ht]
\begin{center}
\includegraphics[scale = 0.55]{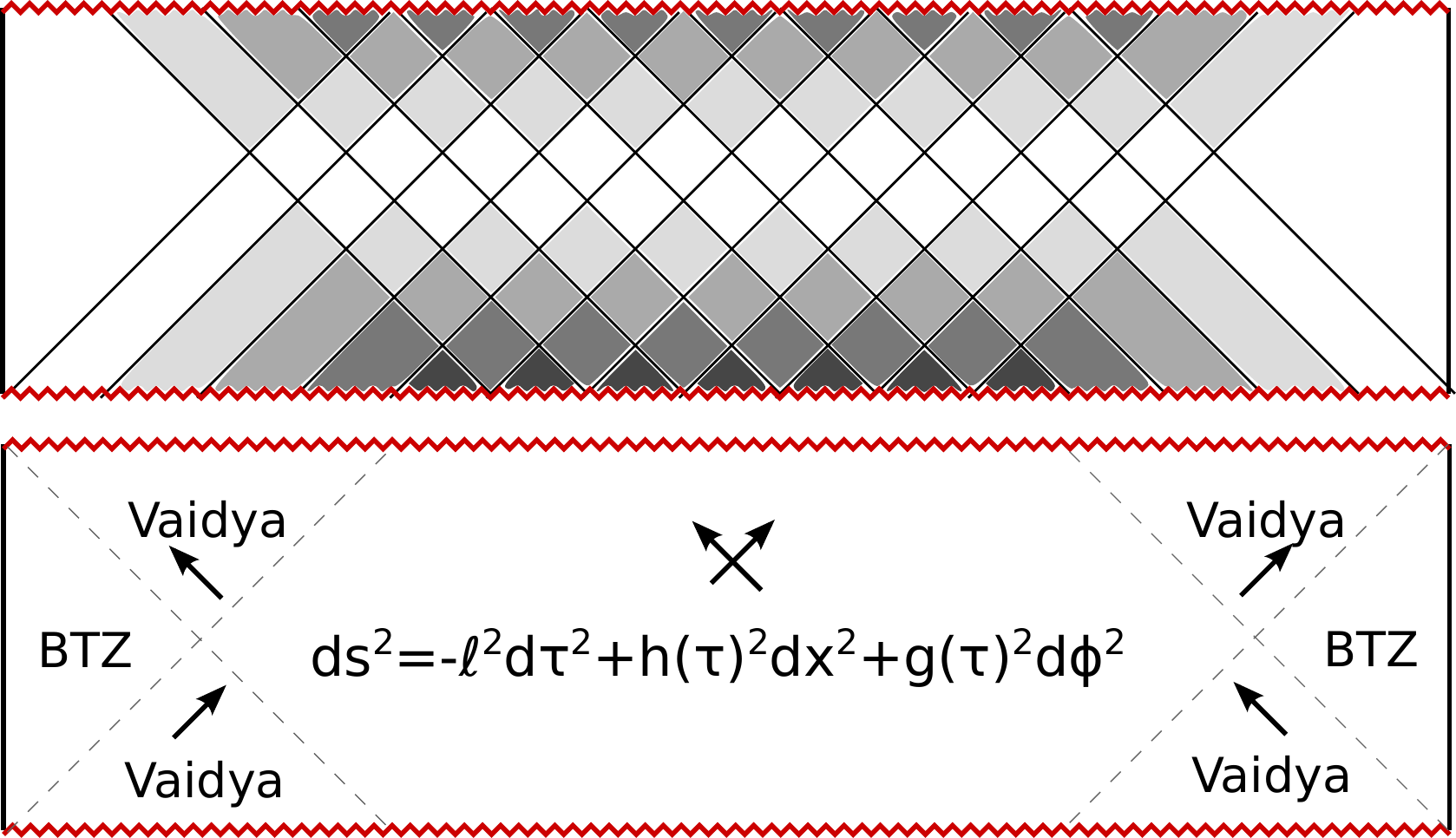}
\caption{The wormhole created from a large number of weak shocks (top) becomes a smooth geometry in the $\alpha\rightarrow 0$ limit (bottom).}\label{figsix}
\end{center}
\end{figure}
In order to check that this metric actually corresponds to the small $\alpha$ limit of the dense network of shock waves, we write down recursion relations for the patched-together geometry in Appendix \ref{recursion}. By taking $\alpha = 0.01$, solving the recursion relations numerically, and computing the size of the $S^1$ as a function of proper time in the direction orthogonal to the symmetry axis, we find excellent agreement with the function $g(\tau)$.

The metric (\ref{metric}) gives us the translationally invariant part in the interior of the wormhole. To complete the geometry, we need to understand how to patch it together with the BTZ exteriors. Here, we go back to the shock wave construction sketched in Fig.~\ref{figsix}, and notice that the intersecting network of shocks in the interior of the wormhole is matched to the empty exteriors across a region in which the shock waves are moving in only one direction. These regions are therefore a piece of the BTZ-Vaidya spacetime, with mass profile determined in Appendix \ref{match}.

\section{Ensembles}\label{ensembles}
In the previous section we have discussed a family of geometries with long wormholes, describing  weak  correlation between the left and right CFTs.   In particular, by taking a large number of shocks or large time separations, the wormhole length can exceed $S$, consistent with a two point correlator of order $e^{-S}$, the value in a typical state found by Marolf and Polchinski \cite{Marolf:2013dba}. However, as we will emphasize in the Discussion, the states constructed in this manner are not typical in the two-CFT Hilbert space.

In this section, we will put the $W$ states aside and address the question of whether truly typical states could be described by smooth geometries. First let us define ``typical state" more carefully.  This concept is straightforward in classical statistical mechanics.  The standard phase space measure on an energy shell in phase space determines the probability for finding a phase space region.   Typical regions are those with typical probability in this measure.  
For an ergodic system time evolution reproduces this probability.   The fraction of time such a system spends in a region is equal to the measure of the region.  So typical states can also be defined as ones that occur typically in the time evolution of the system.

Quantum mechanics is different. If a state $|\psi\rangle = \sum_s  c_s |E_s\rangle$ then
\be
 |\psi(t)\rangle = \sum_s c_se^{-iE_st} |E_s\rangle  .  
 \ee
 Time evolution does not change the magnitude of the coefficient of an eigenvector, only its phase.  But there are natural notions of a distribution for the magnitudes.  For example, in a Hilbert space of dimension $D$, there is unique distribution that is invariant under $U(D)$ transformations. This is given by acting on a reference state with a Haar random unitary.\footnote{Random matrix techniques show that the eigenstates of a random Hamiltonian are distributed in the same way as states obtained by acting  a random unitary on a reference basis.
 }
For large $D$, the probability is proportional to
\be \label{randomgauss}
P(|\psi\rangle) \sim \exp(-\sum_{s=1}^D |c_s|^2/2f^2) 
\ee
where $f$ is chosen so that the state normalization condition $\langle\psi|\psi\rangle = 1$ is satisfied (up to small fluctuations), $2f^2= 1/D$.  This measure gives a natural notion of a typical state.   In a less completely random situation we expect the probabilities in an ensemble to depend on the energy of states.  A natural generalization of (\ref{randomgauss}) to this case is
\be \label{gauss}
P(|\psi\rangle) \sim \exp(-\sum_{s=1}^D |c_s|^2/2f^2(E_s)) 
\ee
where $f$ is smooth over the spread in energies of the system being sampled, and satisfies the normalization condition $\sum_s 2f(E_s)^2 = 1$. The ensemble (\ref{gauss}) provides a natural, but not unique, notion of a typical state.   Note that this ensemble is invariant under time evolution, which just changes the phases of the $c_s$.

We now turn to the question of how time evolution can approximate this ensemble.   Assuming that the Hamiltonian of the system $H$ is sufficiently chaotic, and that the initial state is typical with respect to this distribution,  then time evolution eventually brings this state to within a distance of order one of nearly all states in the ensemble. To see this, we compute
\begin{align}
\int d|\psi\rangle d|\chi\rangle &P(|\psi\rangle)P(|\chi\rangle)\max_t |\langle\chi|e^{-iHt}|\psi\rangle| \\
&=\frac{1}{\mathcal{N}^2}\int \prod_{s}\left(d^2c_sd^2c_s' e^{-(|c_s|^2+|c_s'|^2)/2f^2(E_s)}\right)\max_t \big|\sum_r c_r^*c_r'e^{-iE_rt}\big| \\
&\approx\sum_s\left(\frac{1}{2\pi f^2(E_s)}\int d^2c_s e^{-|c_s|^2/2f^2(E_s)}|c_s|\right)^2 \\
&=\sum_s \frac{\pi}{2} f^2(E_s)= \frac{\pi}{4}.
\end{align}
In the second equality, we have used the assumption that all energy levels are incommensurate, so we can find a time $t$ such that $c_s^*c_s'e^{-iE_st} = |c_s||c_s'|$ for nearly all $s$ (this time will typically be double-exponential in the entropy $S$). The factor $\mathcal{N}$ normalizes the probability distribution. In the final equality, we used the normalization condition for $f$.

In our specific situation we will imagine following \cite{Marolf:2013dba} and adding a  weak ``wire" between the left and right sides that lets the system as a whole thermalize.   We can imagine the wire allowing the exchange of one quantum with thermal energy between the left and right sides every large number of thermal times.  Denote this wire by an operator $\Omega$ which is a  smeared product of local operators in the left and right systems and the total Hamiltonian $H = H_0 + \Omega$ where $H_0 = H_L +H_R$.    Now thermalize by evolving $|TFD\rangle$ forward with $U(t)$. By choosing a random time $t$, we form an ensemble of states that is invariant under time translation. How similar is this ensemble to (\ref{gauss})? We expect the expansion of $|TFD\rangle$ in eigenstates of $H$ to have coefficients $|c_s|$ that are typical of the distribution (\ref{gauss}) for an appropriate $f(E_s)$. Therefore, after some time the state comes within an overlap of $\pi/4$ of any typical state in that ensemble.\footnote{To improve upon the $\pi/4$, we could take our initial state and evolve it with two different chaotic Hamiltonians (``wires'') for various lengths of time in various orders. To be safe one should use order $D$ different time evolution intervals.} This overlap is enough to ensure that the states cannot be distinguished, with an optimal measurement of a linear operator, with probability better than roughly $80\%$.

The ensemble generated by the wire raises a question of time scales: how much evolution is required to produce a state that we may treat as typical? As a lower bound, it seems reasonable to allow at least a time $S$, so that all quanta can equilibrate across the wire. An (extreme?) upper bound is provided by the quantum recurrence time, schematically $\sim e^{e^S}$. Another potentially interesting time scale is the time $\sim e^{S}$, after which point states can be written as a superposition of naively orthogonal states at earlier times.  These recurrence  timescales, if relevant, would be vastly longer than those over which the geometrical constructions of the previous sections are reliable.

Having defined these ensembles, we will now use their time-translation invariance to derive a constraint.  Suppose that a typical state $|\psi\rangle$ is described by a smooth geometry with a long wormhole. Then $U(-t)|\psi\rangle$ is also typical, and hence by assumption also described by a smooth geometry with a long wormhole. Roughly, the two geometries are related by a boost. This is dangerous: imagine that part of the matter supporting the $|\psi\rangle$ wormhole is a light ray behind the horizon.
\begin{figure}[ht]
\begin{center}
\includegraphics[scale = 0.6]{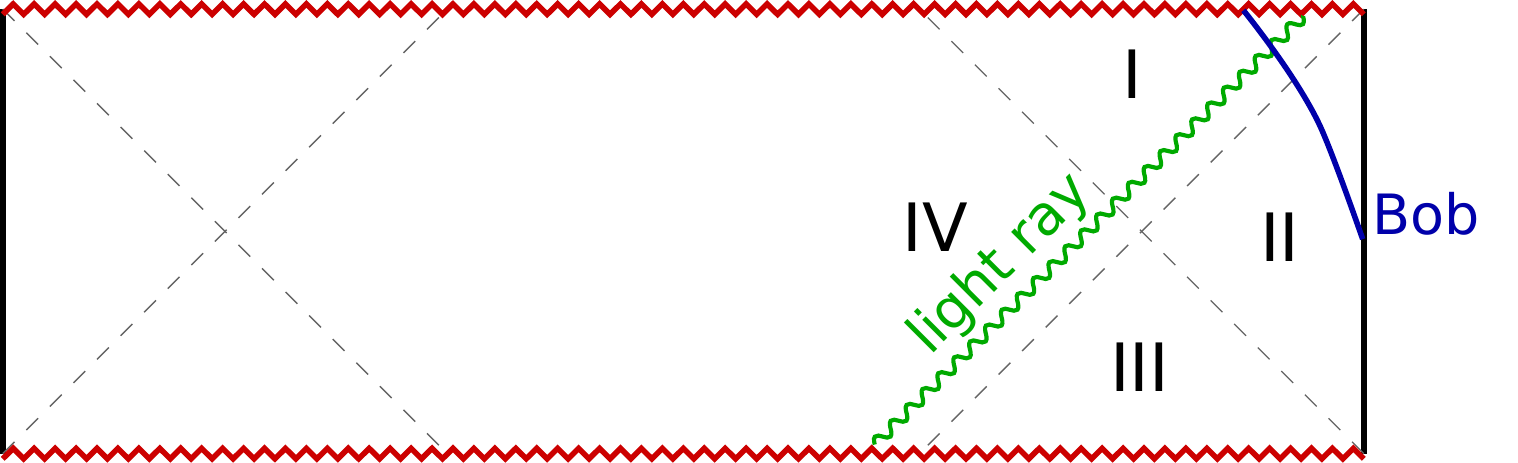}
\caption{Bob falls in from the boundary at $t_B = 0$ and experiences a mild interaction with the stress energy supporting the solution. If he jumped in at a much earlier time $t_B \sim -t_*$, he would experience a dramatic interaction.}\label{fignine}
\end{center}
\end{figure}
If Bob starts falling into the $|\psi\rangle$ black hole at time $t_B=0$, he might experience a mild collision.  But consider the geometry associated with $U(-t)|\psi\rangle$. If Bob falls into this geometry at  time $t_B=0$ his  experience  will be the same  as falling into $|\psi\rangle$ at time $t_B = -t$.  If $t \sim t_*$, Bob will experience a violent collision. 

\begin{figure}[ht]
\begin{center}
\includegraphics[scale = 0.6]{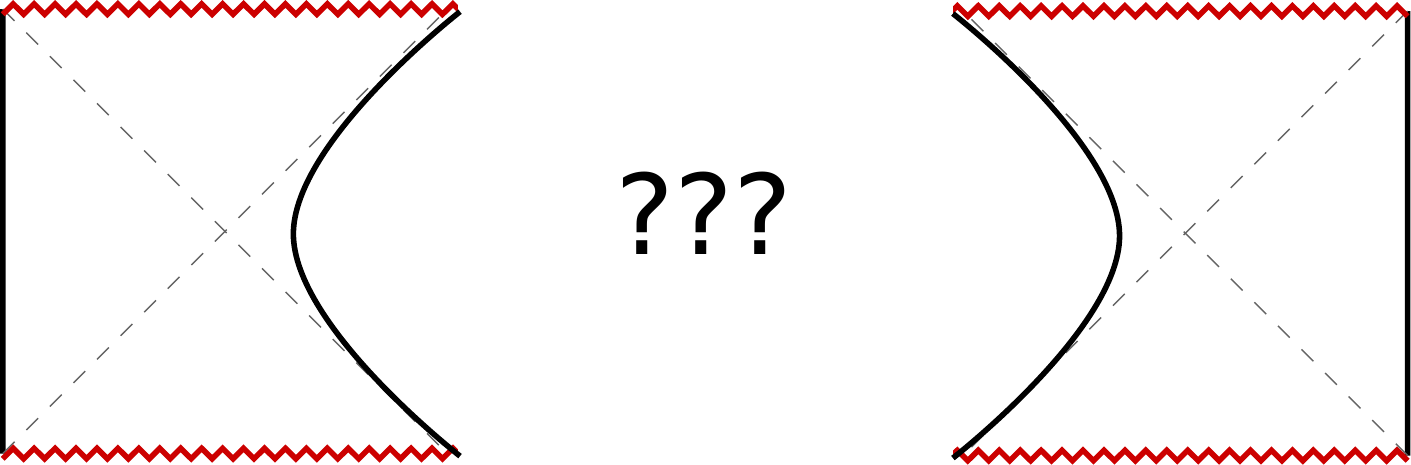}
\caption{A candidate for the geometrical dual to a typical state?}\label{figcandidate}
\end{center}
\end{figure}

It typical states are dual to smooth geometries, avoiding this boosting effect would require all three regions $I, II, III$ on the figure to be essentially the same as the empty eternal black hole. This is a  powerful constraint on the form of such geometries.  These empty regions would have to be joined in some way onto a long wormhole.  The joining locus on the Penrose diagram (Fig.~\ref{figcandidate}) would have to be a surface containing timelike curves of infinite length, quite different from the intuitive notion of a long thin wormhole.
If we imagine this curve to be boost invariant, the configuration in quadrant IV resembles the dual of a cut off CFT.  This suggests that there are other quantum states present than the standard ones at the UV boundary of quadrant $II$.\footnote{The ``mirror operators" of \cite{PR} might be candidates for these. This possibility arose in a discussion with Juan Maldacena and Edward Witten.}

Of course another possibility is that typical states do not have smooth geometries outside of region $II$ \cite{VanRaamsdonk:2013sza}.  An observer falling through the horizon immediately encounters a firewall \cite{AMPS}.

\section{Discussion}\label{discussion}

In the context of 2+1 dimensional Einstein gravity, we have identified a large class of two-sided AdS black hole geometries with long wormholes. These geometries are dual to perturbations of the thermofield double state of two CFTs,
\be
W_n(t_n)...W_1(t_1)|TFD\rangle,\label{hi}
\ee
and they provide constructible examples of highly entangled states with two-sided correlators that are small at all times. The key geometrical effect is boost enhancement of the $G_N$-suppressed backreaction associated to each perturbation \cite{SS}. If the time between perturbations is sufficiently large, their shock wave backreaction must be included, lengthening the wormhole. 

The scrambling time $t_*$ emerges as an important dynamical timescale in the construction of the metrics. For example, perturbations at widely separated times, $\Delta t \sim 2t_*$, create kinked geometries with high energy shocks, while large numbers of perturbations at smaller time separation lead to smoother wormholes. As a second example, even though a multi-$W$ state includes operators local at $n$ different times, if the separations $|t_{i+1} - t_i|$ are greater than $t_*$, our bulk analysis indicated that the CFT state (\ref{hi}) has a locally detectable disturbance only at the ``outermost'' time $t_n$. Roughly, the action of $W_n(t_n)$ disturbs the delicate tuning required for a local perturbation to appear at time $t_{n-1}$; in bulk language, the $W_{n-1}$ shock is captured by a tiny increase in size of the horizon due to the $W_n$ shock.

Although these states display the very small correlation between L and R characteristic of typical states, they are atypical in important ways. 
They have a distinguished time, $t_n$, at which a shock wave approaches the boundary. Also, the $W$ operators increase the energy without increasing the two-sided entanglement. In a typical ensemble, the distribution of entanglement is very sharply peaked, and deficits are highly suppressed in the measure \cite{aspectsofgenericentanglement}. Another feature of these states is that boosting them gives a high energy shock wave on the horizon.  If typical states are dual to smooth geometries, they would have to be of the kind discussed  in \S\ref{ensembles}.

One could attempt to build a typical state out  of  a basis consisting of  the multi $W$ states, each described by a geometry.  It might seem unlikely that a superposition of distinct geometries could again be represented as a geometry, but this is difficult to exclude: in expectation values, the large number of off diagonal terms will dominate, rendering semiclassical reasoning invalid.

By estimating correlators using geodesic distance, we have ignored the backreaction of the field sourced by the correlated operators. Although this should provide an upper bound on the correlation, an interesting possibility is that nonlinear effects might make it possible for relatively short wormholes with high energy shocks running between the singularities to represent states with $\sim e^{-S}$ local correlation between the two sides.

Using the methods discussed in this paper it is straightforward to construct states containing a few particles behind the horizon.   Constructing actual field operators in this region is an open and interesting problem.

\section*{Acknowledgements}
We are grateful to Raphael Bousso, Persi Diaconis, Patrick Hayden, Stefan Leichenauer, Juan Maldacena, Don Marolf, Joe Polchinski, Lenny Susskind, and Edward Witten for discussions.  This  work is supported in part by NSF Grant 0756174.
\begin{appendix}
\section{Recursion relations for many shock waves}\label{recursion}
In this appendix, we will write the recursion relations for the translationally-invariant network of intersecting shock waves. By solving these relations numerically in the $\alpha\rightarrow 0$ limit, one finds agreement with the smooth metric given in Eq.~(\ref{metric}).

\begin{figure}[ht]
\begin{center}
\includegraphics[scale = 0.75]{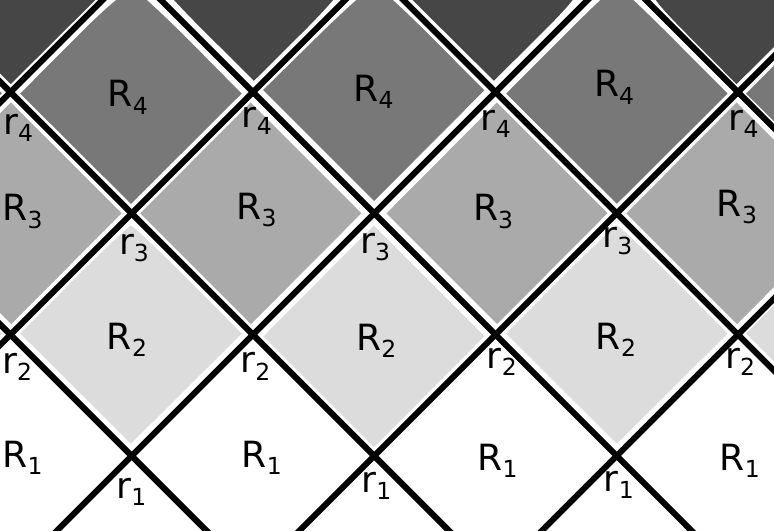}
\caption{The size of the $S^1$ at the vertices is labeled $r_n$, and the $R$ parameter of the BTZ geometry forming each plaquette is labeled $R_n$.}\label{figbtz}
\end{center}
\end{figure}

Exploiting the discrete translational invariance of the arrangement of shock waves, we can represent the metric in terms of the radii of the collisions, $\{r_n\}$, and the BTZ $R$ parameters of the geometries between collisions, $\{R_n\}$ (see Fig.~\ref{figbtz}). We would like to check the function $g(\tau)$ in the case $\ell = R = 1$. In order to do so, we will write recursion relations for $r_n$ and $R_n$, and then compute the geodesic distance ``straight up'' from the first collision to the $n$'th. Identifying this with the interval in $\tau$, we will then be able to confirm that the radius of the $S^1$ (determined by $r_n$) depends on $\tau$ as $g(\tau)$.

We need two recursion relations, one each for $r_n$ and $R_n$. One of these equations is given simply by applying the DTR relation Eq.~(\ref{DTR}) at a given vertex, with $f(r) = r^2 - R_n^2$. This gives
\be
R^2_{n+1} = r^2_n  + \frac{(R^2_{n} - r^2_n)^2}{R^2_{n-1} - r_n^2}.\label{rone}
\ee
To get the other equation, we proceed as follows. We focus on a given plaquette, with BTZ parameter $R_n$, and assume that we know the radii $r_n,r_{n-1}$ of the side and bottom vertices. Let us choose a Kruskal frame for this patch in which $u = v = u_b$ at the bottom vertex. Then using Eq.~(\ref{ruv}) we must have
\be
\frac{r_{n-1}}{R_n} = \frac{1 - u_b^2}{1+u_b^2}.
\ee
Now, holding $v = u_b$ fixed, we solve for $\Delta$, the change in $u$ that is necessary to reach the radius of the side vertex, $r_n$. The radius of the top vertex is then determined by 
\be
\frac{r_{n+1}}{R_{n}} = \frac{1 - (u_b+\Delta)^2}{1+(u_b+\Delta)^2}.
\ee
Eliminating $u_b$ and $\Delta$, we find the recursion relation
\be
r_{n+1} = \frac{2r_{n}R^2_{n} - r_{n-1}R^2_{n} - r_{n-1}r_{n}^2}{R^2_n + r_n^2 - 2r_nr_{n-1}}.\label{rtwo}
\ee

For a wormhole that connects BTZ regions with $R = 1$, the initial conditions are $R_1 = r_1 = 1$. Since the recursion relations are second order, we also need to determine $R_2$ and $r_2$. These can be found using the two-shock solution:
\be
r_2 = \frac{1 - \alpha^2}{1+\alpha^2}, \hspace{20pt} R_2 = \sqrt{1 + 4\alpha^2}.
\ee
The equations (\ref{rone}) and (\ref{rtwo}), together with these initial conditions, completely determine the geometry. In order to compare with the smooth wormhole, we also need to compute the geodesic distance ``straight upwards.'' Using $u_b$ and $\Delta$ derived above, along with the Kruskal metric Eq.~(\ref{kruskal}), one can check that the timelike distance from the bottom vertex to the top vertex of the $n$'th plaquette is
\be
2\tan^{-1}\sqrt{\frac{R_n + r_{n-1}}{R_n - r_{n-1}}} \frac{R-r_n}{R+r_n}- 2\tan^{-1}\sqrt{\frac{R_n - r_{n-1}}{R_n+r_{n-1}}}.
\ee
Taking $\alpha = 0.01$, numerically solving the recursion relations, and plotting $r_n$ as a function of the total geodesic distance from the initial slice, one finds excellent agreement with $g(\tau)$.

\section{Vaidya matching conditions}\label{match}
We will work out the matching condition in detail for the top left Vaidya region in the lower panel of Fig.~\ref{figsix}. This is a portion of the geometry
\be
ds^2 = (\rho(V)^2-r^2)dV^2+2\ell drdV + r^2d\phi^2.
\ee
The $V$ coordinate is $-\infty$ on the horizon, and it increases in the inward null direction (i.e. up and to the right). The function $\rho(V)$ is determined by matching onto the metric in Eq.~(\ref{metric}) across a null slice. In particular, we require that the metric should be $C^1$ across the matching surface.\footnote{Requiring continuity alone would allow a $\delta$-function stress tensor traveling along the null surface.} Continuity of the $S^1$ implies that $r = g(\tau)$ along the join. By taking the derivative along the patching surface, we can relate the normalization between the inward-pointing null vectors in the two coordinate systems. In this way, one finds that $2\ell g'(\tau) d\tau = (r^2 - \rho^2(V))dV$ along the surface. The $C^1$ property of the metric relates the normalization of the outward-pointing null vectors, by matching the derivative of the size of the $S^1$. Requiring the inner product of these vectors to be continuous across the matching surface, we find $g'(\tau)^2 = \rho^2(V) - r^2$. Rearranging these equations, we determine $\rho(V)$ as follows. First, find $V(\tau)$ along the matching surface via
\be
V(\tau) = -2\ell\int^\tau\frac{d\tau}{g'(\tau)}.
\ee
Next, invert this to find $\tau(V)$, and fix $\rho(V)$ using
\be
\rho(V)^2 = g(\tau(V))^2 + g'(\tau(V))^2.
\ee
For our specific $g(\tau)$, we were not able to compute $\rho(V)$ exactly.\footnote{A surprisingly good approximation to the metric is $g(\tau) \approx R -R\tau^2$, from which one finds $\rho(V) \approx R + Re^{2RV/\ell}$.} However, it is clear that these conditions completely fix the geometry, up to the undetermined overall length of the central region of the wormhole.

\end{appendix}

\end{document}